# Liquid Crystals as Multifunctional Interfaces for Trapping and Characterizing Microplastics

Fiona Mukherjee[1,2], Anye Shi[1], Xin Wang[1], Fengqi You[1]*, and Nicholas L. Abbott[1,2]*

[1] Smith School of Chemical and Biomolecular Engineering, Cornell University, Ithaca, NY 14853, USA

[2] Department of Chemistry and Chemical Biology, Cornell University, Ithaca, NY 14853, USA

**ABSTRACT:** Identifying and removing microplastics (MPs) from the environment is a global challenge. This study explores how the colloidal fraction of common MPs behave at aqueous interfaces of liquid crystal (LC) films. We observed polyethylene (PE) and polystyrene (PS) microparticles to be captured at the LC interface and exhibit distinct two-dimensional aggregation patterns. The addition of low concentrations of surfactant (sodium dodecylsulfate (SDS)) was found to further amplify the differences in PS/PE aggregation patterns, with PS changing from a linear chain-like morphology to a singly dispersed state with increasing SDS concentration and PE forming dense clusters at all SDS concentrations. Statistical characterization of assembly patterns using fractal geometric theory-based machine learning and a deep learning image recognition model yielded highly accurate classification of PE vs PS (>99%). Additionally, by performing feature importance analysis on our deep learning model, dense, multi-branched assemblies were confirmed to be unique features of PE relative to PS. To obtain additional insight into the origin of these key features, we performed microscopic characterization of LC ordering at the microparticle surfaces. These observations led us to predict that both microparticle types should generate LC-mediated interactions (due to elastic strain) with a dipolar symmetry, a prediction consistent with the observed interfacial organization of PS but not PE. We conclude that the non-equilibrium organization of the PE microparticles arises from their polycrystalline nature, which leads to rough particle surfaces and weakened LC elastic interactions and enhanced capillary forces. Overall, our results highlight the potential utility of LC interfaces for surface-sensitive characterization of colloidal MPs.

INTRODUCTION

Microplastic (MP) is a term that has been adopted to describe sub-millimeter-sized plastic debris that is either fragmented from bigger plastic waste or synthesized as additives in commercial personal-care products. Common polymers found as MPs are polyethylene (PE), polystyrene (PS), and polymethylmethacrylate. MPs are potential carriers of toxic organic and heavy metal components, are ingested by marine animals and fish[1–6] and end up localized in their guts, and are also known to penetrate human tissues more easily than larger particles.[7–9] While the health risks and environmental consequences of microplastics appear potentially significant, capture and identification of MPs in the environment is an unresolved challenge due to their heterogeneity in chemical composition, shape and size.[10–14]

The surfaces of MPs are important because surface characteristics such as surface chemical composition, charge or roughness determine colloidal interactions of MPs with the environment,[15] which impacts their position in the water column,[16–20] their propensity to adsorb organic compounds[3,21] their uptake by marine organisms,[1,2,22] and their tendency to coagulate.[23–26] A fundamental challenge with the characterization of MPs is that their interfaces change over time in the environment.[12] The most widely used methods for MP characterization, specifically Raman[27] and Fourier Transform Infrared (FTIR) spectroscopy,[28–31] however, are not surface-selective and generate spectra that are dominated by bulk properties of MP particles.[32] Accordingly, the development of surface-sensitive methods for characterization of MPs has the potential to be useful for understanding and addressing the MP problem.[13]

In this paper, we set out to explore how the colloidal fraction of two common MPs, PS and PE, behave at liquid crystal (LC)-aqueous interfaces with the long-term goal of enabling new approaches to MP characterization that are surface-sensitive. Our study goes beyond past investigations of interactions between microparticles at isotropic oil/water interfaces[33–39] (e.g. Pickering emulsions)[40–43] in that we use anisotropic oils (i.e., LCs) to characterize MP because the organization of microparticles in LCs[44–47] and at LC interfaces is particularly sensitive to the surface properties of microparticles.[48–53] This surface-sensitivity arises from the ordering of LCs induced by surface interactions, the so-called surface anchoring of LCs. Several past studies have explored the formation of assemblies by colloidal and molecular adsorbates at aqueous interfaces of nematic LCs, including with amphiphiles present on the interface (e.g., dodecyltrimethylammonium bromide (DTAB), sodium dodecyl sulfate (SDS), or 1,2-dilauroyl-sn-glycero-3-phosphocholine (DLPC)).[54] These prior studies establish that microparticles adsorbed at the LC interface perturb the ordering of the LC, which costs elastic energy and introduces topological defects, and thereby generates both short (due to defects) and long range (due to elastic energies) interactions between the microparticles.

We chose PS and PE microparticles as model MPs in our study because of their abundance in marine plastic contaminants.[2–6,55–57] Additionally, it has been previously reported that nematic LCs anchor differently on PS and PE surfaces, assuming a tangential orientation at the surfaces of PS microparticles and a radial orientation on the surfaces of PE microparticles.[58,59] Our finding that PS and PE microparticles form assemblies at LC-aqueous interfaces with unique organizations generated a number of questions that we address in this paper. 1. How are the organizational patterns exhibited by the microparticles influenced by LC ordering at the aqueous-LC interface? To address this question, we vary the LC anchoring at the aqueous interface of the LC via the addition of surfactant and study its effect on microparticle organization. 2. Can we use the patterns formed by microparticles at LC interfaces to determine the identity of the polymer within the microparticles? Although inspection of optical images of the LC interface reported in this paper suggest that PE and PS microparticles do form distinct patterns at the LC interface, we also encountered substantial sample-to-sample variation. To address the issue of accuracy of classification, we provide a statistical analysis of assembly patterns using two classification strategies; (i) fractal dimension-based machine-learning, a geometry-based classification approach, and (ii) a novel image-based convolution neural network (CNN). We analyse the importance of aggregation pattern features used by the CNN to permit PE to be distinguished from PS with high accuracy. 3. Finally, we explore the microscopic organization of LC around the PE and PS microparticles, and ask if the assembly patterns characterized by the CNN can be understood from equilibrium considerations. Overall, the answers to these questions provide support for our proposal that the surface-sensitivity of LC interfacial ordering offers the basis of new ideas and methods for the capture and characterization of MP microparticles.

MATERIALS AND METHODS

**Materials:** Chemicals were used as purchased from manufacturers without further purification. 5CB (>99.5%) was purchased from Jiangsu Hecheng Advanced Materials Co. Ltd. (Nanjing, China). NaCl (>99%) was purchased from Sigma Aldrich. Fluorescent PS microspheres (FluoSpheres; sulfate-functionalized, 1 and 2μm diameter, yellow-green fluorescent (lex= 505nm, lem= 515nm))was purchased from Thermo Fischer Scientific(Waltham, MA). The 14-20μm polystyrene, 1-4 μm and 10-27 μm polyethylene microparticles (CPMS) were purchased from Cospheric. Deionization of a distilled water source was performed using a Milli-Q system (Millipore, Bedford, MA). 20μm-thick copper transmission electron microscopy (TEM) grids were purchased from Electron Microscopy Sciences.

**Preparation of Octyltrichlorosilane-Coated Glass Slides:** Glass microscope slides (Fisher Scientific) cleaned with Alconox detergent, ethanol (Fisher Scientific) and deionized (DI) water were incubated in Nochromix solution (a mixture of Nochromix and 98% sulfuric acid) for 24 h. The slides were thoroughly rinsed with DI water followed by sonication in water for 5 min. This was followed by a final washing step with DI water and ethanol before drying under a stream of gaseous $N_2$. The clean glass slides were subsequently incubated in a glass staining jar with 1 vol% octyltrichlorosilane (OTS, Sigma-Aldrich) in hexane (Fisher Scientific) for 30 min. After silanization, the glass slides were washed sequentially with hexane (100 mL), chloroform (100 mL × 2), and water (100 mL).

**Fabrication of Millifluidic Sensing Device:** Figure 1a shows a schematic illustration of the millifluidic channel used in our experiments. The millifluidic channel was composed of three layers. The top two layers were made of acrylic sheet (1.5 mm thickness from ZLazr.com, USA), laser-cut into uniform dimension (75x26 mm) of a standard glass slide, and equipped with inlet and outlet ports for sample introduction. The bottom layer was an OTS-treated glass microscope slide as described above. The two top layers were permanently bonded by applying chloroform to the contacting surfaces of the acrylic sheets and rested overnight. This was followed by a thorough cleaning step with sequential immersion of the acrylic sheet layers in water, ethanol and hexane. Residual solvents were dried off under a steam of gaseous $N_2$.

**Preparation of Thin LC Films:** The LC used in our study was 4-cyano-4'-pentylbiphenyl (5CB), a thermotropic LC that forms a nematic LC between 24 and 35 °C. 20μm-thick LC films were prepared by pipetting 0.2μL of 5CB into the pores of 75 mesh (thickness 20μm; lateral pore size 284μm) transmission electron microscopy (TEM) grids that were supported on OTS-treated glass slides. Excess LC was then removed by a micropipette tip using capillary action to produce a flat LC film with a thickness of 20μm.

**Optical Microscopy:** The imaging was carried out using an Olympus BX41 microscope equipped with 4×, 20×, 50x objectives and two rotating polarizers. A Moticam 10.0 MP camera was used for optical microscopy.

**Preparation of Microparticle Solution:** The 1-2 μm PS microparticles were cleaned by centrifuging for 5 minutes, a 0.01 wt% microparticle solution at 3000 rpm, three times. This process was performed to exclude physisorbed surfactants and stabilizers from the PS microparticle surface so that the surface was chemically similar to the bigger 14-20 μm PS microparticles which were purchased dry, and without added stabilizers (from Cospheric, CPMS). Following this, a measured concentration of microparticles were dispersed in aqueous solutions of 0.3M NaCl and SDS at various concentrations. The PE microparticles were used as purchased and due to a similar synthetic procedure, the 1-4μm and 20-27μm PE microparticles were assumed to have similar surface properties. Due to hydrophobicity of the PE microparticles, it was difficult to disperse them in water. However, with sonication for 20 minutes and vortexing at 3000 rpm for 5 minutes, the microparticles could be singly dispersed in aqueous solution and were introduced into the millifluidic channel through the inlet using a glass pipette.

**Zeta Potential Measurement:** PS and PE microparticles were dispersed in different concentrations of electrolyte NaCl, LC-in-water emulsions were formed by homogenizing a mixture of 5CB in aqueous solutions comprising either water only or in specific concentrations of electrolyte for 30 s. After allowing 30 min for the sample to equilibrate, $\zeta$ on the aqueous side of the LC-aqueous interface was measured using the Malvern Zetasizer Nano instrument.

**Microplastic Image Dataset (MP dataset):** We collected brightfield optical microscope images of PS and PE microparticle aggregation patterns at 5CB-aqueous interfaces supported on TEM grids. We varied the concentration of the aqueous phase as 0.01mM, 0.025mM, and 0.1mM SDS, in the presence of 0.3M NaCl. To form a microplastic image dataset (MP image dataset) with 355 images of PS and 491 images of PE aggregation patterns, we performed 90 independent experiments. For convenience, PS10, PS25, and PS100 mean PS in 0.01mM, 0.025mM, and 0.1mM SDS, respectively. PE10,

PE25, and PE100 indicate PE in 0.01mM, 0.025mM, and 0.1mM SDS. This dataset is the main source for deep learning model training, feature extraction-based machine learning study, and GradCAM visualization.

**Fractal Dimension Analysis:** We used a self-designed AI-aided FD calculation algorithm to compute FD values over the largest cluster of each image. The reason to compute the largest cluster's FD, instead of the average FD value for all the clusters, was the presence of dispersed microparticles outside of the aggregate structure for some images in MP image dataset. Dispersed microparticles have extremely low FD values (around 1.0 -1.04), as a result of which, the average FD values can decrease drastically for images with lots of singly dispersed microparticles. The largest cluster reflects the assembly preference of a particular microparticle type, which we sought to use as a metric to quantify microparticle aggregation patterns.

The FD calculation algorithm has three steps: cluster segmentation, cluster screening and box-counting for FD computation. The input RGB images are first converted to binary images to segment the clusters' boundaries. The perimeter of each cluster is counted and ranked to screen out the largest cluster. A list of box sizes in pixels, s = [2, 4, 6, 8, 16], is used to surround the largest cluster. The number of boxes, N(s), is optimized via a mixed integer optimization algorithm[60]. The final FD value is defined as $f = -\frac{\log(N(s))}{\log(s)}$.

We calculated the FD values for all the images in the MP image dataset. The FD results are analysed by computing the mean and standard deviation of each sample category.

**Machine Learning Model Training over Fractal Dimension Dataset:** The machine learning (ML) model training is performed using *Scikit-learn* library in Python. Decision tree, random forest, support vector machine, k-nearest neighbours and naïve Bayesian classifiers are trained directly over MP image dataset. Accuracy, precision, F1, and recall are computed as evaluation metrics. All the machine learning models are trained with 5-folded cross-validation. The machine learning models' hyperparameters are optimized and shown in Table S1-2.

**Deep Learning Model Training over MP Dataset:** The deep learning model training work is implemented using *Keras* library in Python. A more detailed discussion of CNN model functionality can be found in the *Supporting Information* file. The model training steps include: (1) Data pre-processing and augmentation: all the images in MP image dataset are first normalized to RGB formatting images with fixed size (400, 400, 3), representing (height, width, channel). Each image pixel is displayed via three digits from 0 to 255 to show the color scale in R (red), G (green), and B (blue) channels. MP dataset has two classes of data: PS (355 images) and PE (491images), which is randomly stratified with a ratio of 72:18:10 for training, validation, and test data. 5-fold cross-validation and data augmentation are performed to increase model generalization ability. (2) Model training and comparison: we apply multiple baseline CNN models, like VGG16, InceptionV3, Xception, ResNet50, and EfficientNetB0 to train these CNN models over MP dataset in transfer learning styles. Specifically, we borrow the convolutional layers kernel parameters from "*ImageNet*" pretrained models. We freeze the model parameters for the first several convolutional blocks keeping only the last convolutional block and the classification engine open for training. The classification engine sizes are modified to fit the binary classification mode. For instance, pristine VGG16 has 4096, 4096 and 1000 for the last three dense layer sizes. We change the dense sizes to 1024, 1024 and 2 when we perform model training over MP image dataset. The parameters of the last convolutional block and classification engines are updated iteratively to minimize the binary cross entropy loss via Adam optimizer at $10^{-4}$ learning rates. A categorical labelling method is used to label PS/PE as [0,1] and [1,0] respectively. Computers equipped with RTX 3080 GPU served as computation platforms. We performed 100 epochs of iterative training for model accuracy comparison. The best model parameters with the lowest validation loss was saved during iterative parameter updating. The CNN model classification accuracy was evaluated over the whole MP dataset after model training.

*PolyNet* **and Feature Extraction-based Machine Learning Model Training over MP Dataset:** This work is performed using *Keras* and *Scikit-learn* libraries in Python. *PolyNet* model training is performed over feature maps generated via well-trained VGG16 models, which we obtained in the above section. In *PolyNet* training, we retrain a flatten layer and two dense layers with sizes 256 & 2, and activation functions ReLU & softmax, respectively. The *PolyNet* model training follows the same procedure as deep learning model training in the previous section.

Feature extraction-based machine learning models are trained over feature vectors obtained from well-trained VGG16 models after flatten layer process. Each image can be translated into feature vectors (1D data arrays). Five machine learning classifiers (Decision tree, random forest, support vector machine, logistic regression, and naïve Bayesian) are trained to map 1D feature vectors to image labels. Prediction accuracy and mean squared errors (MSE) are computed over the whole MP image dataset.

**Gradient Weighted Class Activation Mapping (GradCAM) Analysis over *PolyNet*:** The feature importance of *PolyNet* is analysed via GradCAM algorithm. We first concatenate well-trained VGG16 convolutional structures and the newly trained *PolyNet* classifier with flatten layer and two dense layers, to form a complete CNN architecture. GradCAM calculates the weights for the feature maps via the gradient $\frac{\partial L^c}{\partial f_k}$, for class loss $L^c$ over feature map $f_k$. The overall GradCAM formula is:

$$\text{GradCAM}_c = \text{ReLU}\left(\sum_k \sum_i \sum_j \frac{\partial L^c}{\partial f_{i,j}^k} f_k(X,Y)\right) \quad [1]$$

$f_k(X,Y)$) represents the *k*-th feature map in the last convolution layer at the spatial location (X, Y), where *i* and *j* denote the spatial location of gradient in feature map $f_k$. ReLU activation is applied to leave non-negative values for normalization and visualization.[61] We plot the GradCAM output in "*jet*" colormap using *matplotlib* in Python to show heatmap value from 0 (blue) to 1 (red).

RESULTS AND DISCUSSION

**Assemblies formed spontaneously by PS and PE microparticles at LC-aqueous interfaces:** Our first goal was to find experimental conditions that would allow us to collect PS and PE microparticles at the LC-aqueous interface. Our experimental setup comprised a millifluidic channel that was used to deliver an aqueous dispersion of microparticles to a 20μm-thick film of the nematic LC 4-cyano-4'-pentylbiphenyl (5CB) (details in Figure S1a) supported within a metal grid (Figure 1a) on a glass substrate. Each square of the grid was 284 x 284 μm², which is large compared to the size of the microparticles used in the majority of our experiments (diameters of 1-4 μm). The supporting glass substrate was coated with octyltrichlorosilane (OTS) to induce a perpendicular alignment of the LC.[54] Because PS microparticles are denser (1.05 g/cm³) than water whereas PE microparticles are less dense (0.96 g/cm³) than water, in the study reported in this paper, we collected PS microparticles by sedimentation downward onto the LC interface and PE microparticles by floatation upward onto the LC interface.

Our initial experiments used microparticles of PS or PE dispersed in pure water. However, brightfield microscopy revealed that neither PE or PS microparticles were adsorbed at the LC-water interface after 24hrs of incubation. We measured the zeta potentials of the PS microparticles in water, PE microparticles in water and the 5CB-water interface (using LC-in-water emulsion droplets) to be $-55 \pm 5$mV, $-23 \pm 6$mV and, $-45 \pm 3$mV, respectively [details in Figure S4]. This led us to propose that electrostatic interactions were likely creating a barrier to adsorption of the microparticles at the LC-aqueous interface.[62,63] Consistent with this conclusion, addition of 0.3 M NaCl to the aqueous phase initiated irreversible adsorption of microparticles at LC-aqueous interface (Figure S5). We found also that addition of a low concentration of surfactant (e.g., 0.025 mM, sodium dodecylsulfate (SDS)) minimized aggregation of the microparticles in the bulk aqueous dispersion (prior to adsorption on the LC interface). The total number of microparticles adsorbed onto the LC interface was observed to increase with time (Figure S5). During an incubation time of 1h, we measured microparticles to be collected at a rate of ~100 microparticles/mm²/min to a final surface concentration of $6000 \pm 600$ microparticles/mm² (Figure 1b,c and details in Figure S5a,b). Significantly, inspection of the brightfield microscope images in Figure 1b,c reveals that the PS microparticles assembled into chain-like assemblies, while PE microparticles formed small and dense aggregates (SDS concentration of 0.025 mM). This first observation hints that PS and PE microparticles form distinct assembly patterns at the LC-aqueous interface. Next, we sought to determine if we could amplify differences observed between the microparticle organization patterns by changing the SDS concentration. Additionally, we investigated if the differences in assembly patterns could be understood on the basis of the LC ordering near the microparticle surfaces and equilibrium considerations.

**Can we amplify differences in the organizational patterns exhibited by PE and PS microparticles at LC interfaces?** With the goal of amplifying differences in the organizational patterns between the PS and PE microparticles at the LC interface, and motivated by past studies of surfactant adsorption at LC-aqueous interfaces,[64–66] we explored the effects of SDS concentration in the bulk aqueous solution. Specifically, past studies have demonstrated that addition of SDS to the aqueous phase causes the LC orientation at an aqueous interface to change from a parallel (planar anchoring) to a perpendicular (homeotropic anchoring)[54,67,68] orientation with increasing SDS concentration (Figure S2).

In addition, in the presence of high concentrations of salt (as used in the experiments reported in this paper), it has been observed that SDS[68–70] at intermediate concentrations undergoes a lateral phase separation on the LC interface to generate coexisting planar and homeotropic domains (Figure S3). The driving force for the phase separation is elastic energy stored in the initial strained state of the LC, as detailed previously.[50,68,69] Because the elastic energy stored in the LC is strongly dependent on the LC film thickness, and because the LC film thickness varies within the TEM grids used in our experiments,[69,70] in the experiments reported below, we observed sample-to-sample variation (approximately 40% of samples containing 0.025 mM SDS exhibited phase separated domains). Below, we describe the behaviors of PS and PE microparticles for samples with and without phase-separated SDS domains at the LC interface.

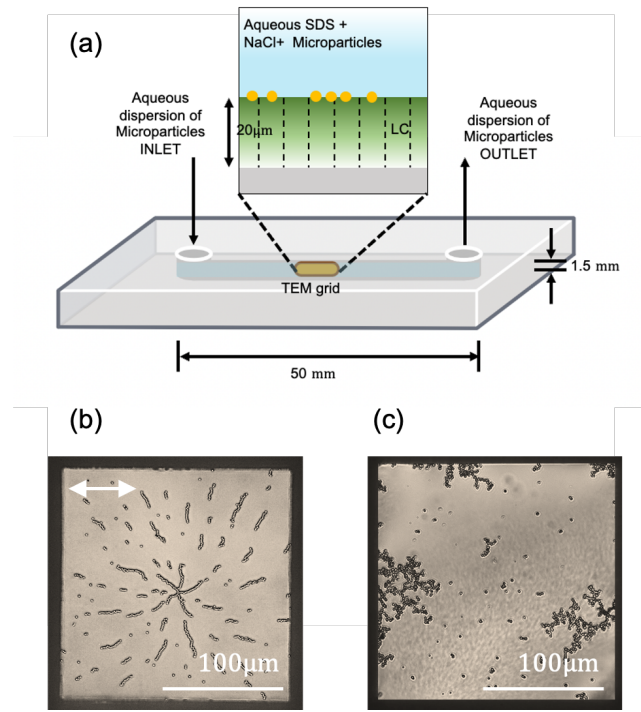

**Figure 1.** (a) Schematic illustration of a millifluidic channel used to obtain the images of microparticles adsorbed to LC-aqueous interfaces. (b and c) Top view of the LC-aqueous interface decorated by PS or PE microparticles (b and c, respectively), as seen under an optical microscope (single polarizer). The aqueous phase from which the microparticles were adsorbed contained 0.3M NaCl and 0.025 mM SDS. The microparticles were adsorbed for 1 h. Scale bar is 100μm.

The brightfield micrographs in Figure 2 (left side) show the change in organization of PS microparticles at the LC interface as we changed the SDS concentration between 0.01 mM, 0.025mM and 0.10 mM (all with 0.3M NaCl). The micrographs on the right side of Figure 2 (crossed polars) reveal the orientation of the LC. For samples containing 0.01mM SDS (Figure 2a), the PS microparticle patterns formed within 1h of incubation were characterized by linear chain-like organizations. We also observed that the linear chains interacted with each other to give the appearance of branching. For samples containing 0.025mM SDS (Figure 2b) that did not exhibit interfacial phase separation of SDS, the PS assemblies again comprised linear chains (Figure 2b) but with fewer branches as compared to 0.01mM SDS (Figure 2a). In contrast, for the phase separated samples, we observed PS to form both short linear chains and singly dispersed microparticles (Figure 2c). Inspection of the polarized light micrograph in Figure 2c reveals that the chains formed within tilted (optically bright) regions of the LC interface, whereas the singly dispersed microparticles formed in the regions of the LC interface with a homeotropic orientation (optically dark). For samples containing 0.1mM SDS, the PS microparticles were singly dispersed on the interface (Figure 2d) and no evidence of chaining was found in any samples. From these experiments, we conclude that the PS microparticles organize into chain-like structures at low SDS concentration (0.01mM), with the extent of aggregation progressively decreasing with increasing SDS concentration (0.025mM, 0.1mM). In contrast to PS, observations shown in Figure 3 (left images) reveal that PE microparticles form small, dense, highly branched clusters on the LC interface across all SDS concentrations investigated. Inspection of the right-side images in Figure 3 confirmed that the orientation of the LC at the aqueous interface did change with SDS concentration, leading us to conclude that variation in the orientation of 5CB at the aqueous interface (using SDS) did not lead to noticeable changes in the organization of PE microparticles. For samples containing 0.025mM SDS, we observed dense, multi-arm clustering of PE microparticles in the absence and presence of interfacial phase separation of SDS (Figure 3b,c) without any obvious influence of LC orientation on microparticle assembly.

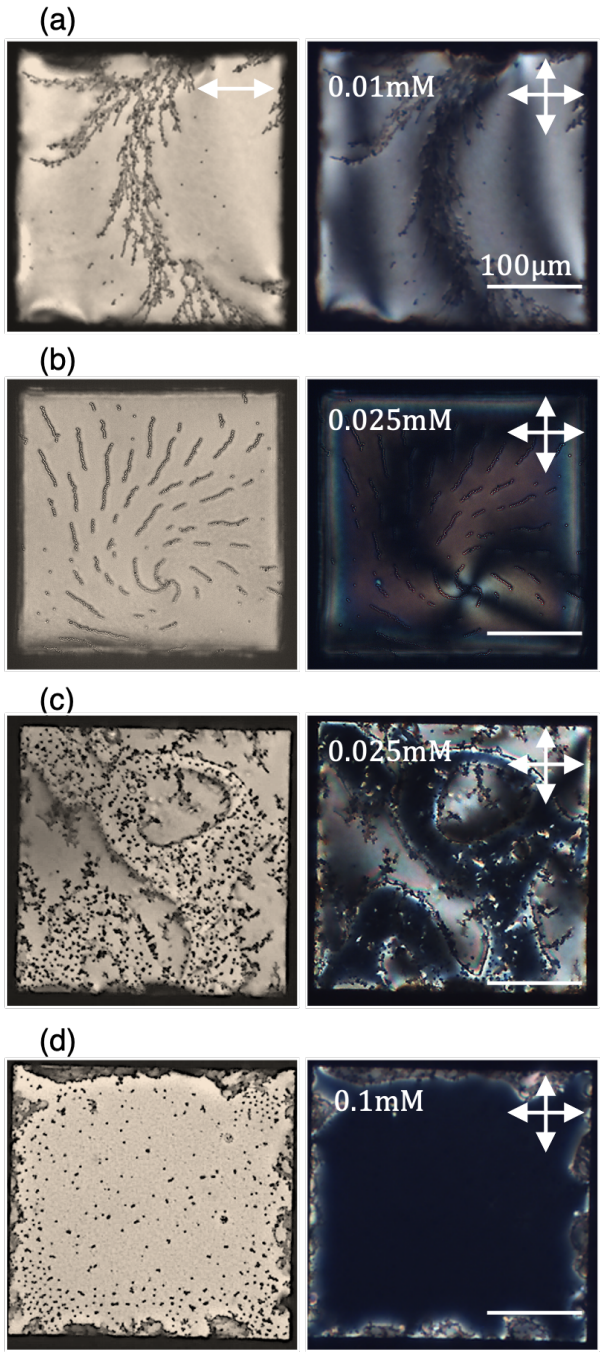

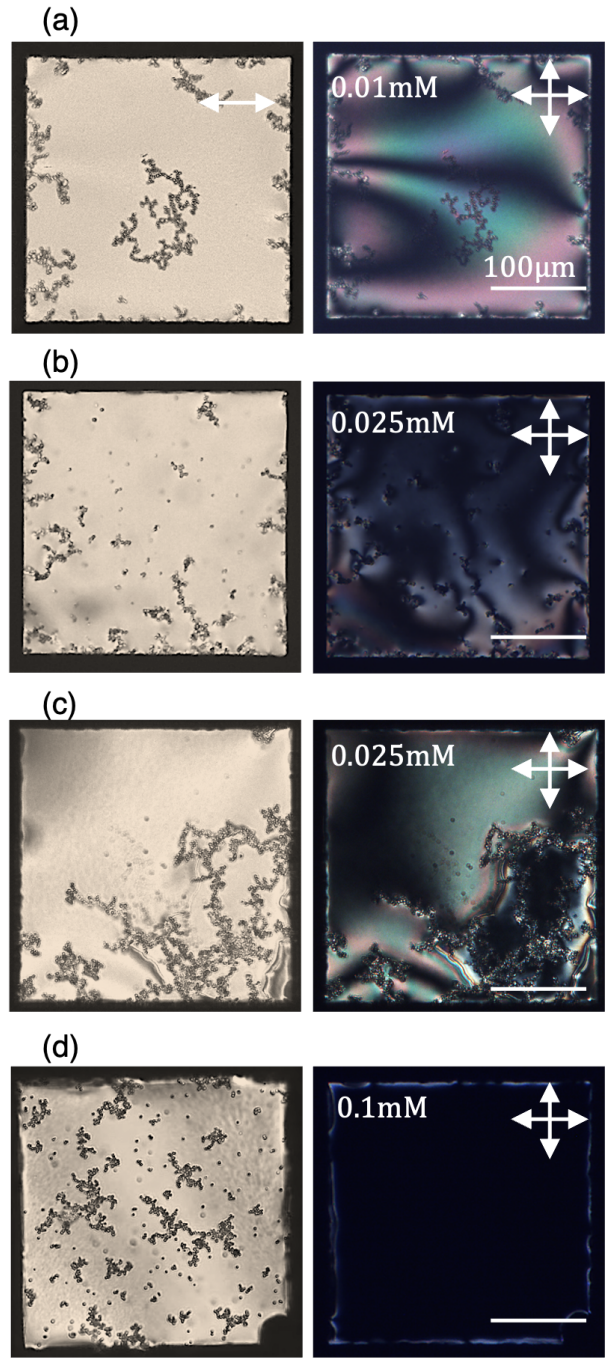

**Figure 2**. (a)-(d) Brightfield (left column) and crossed-polar (right column) optical micrographs showing the organization of 1- 2µm PS microparticles at LC-aqueous interfaces in the presence of 0.01mM, 0.025mM or 0.10 mM SDS and 0.3M NaCl. Scale bar is 100µm.

**Figure 3.** (a)-(d) Brightfield (left column) and cross polar (right column) optical micrographs showing the organization of 1-4µm PE microparticles at LC-aqueous interfaces in the presence of 0.01 mM, 0.025mM or 0.10 mM SDS and 0.3M NaCl. Scale bar is 100µm.

Overall, a comparison of the results shown in Figure 2 and Figure 3 suggests that SDS concentration does impact our ability to distinguish between PE and PS based on their aggregation patterns at the LC interface. In particular, we observe that at 0.1 mM SDS concentration, the PS microparticles trapped at the LC interface are singly dispersed whereas PE microparticles aggregate into dense, fractal-like clusters. However, we also observed significant variation in the assembly patterns across multiple independent experiments (Figure S8-2). This sample-to-sample variation motivated us to explore statistical methods to classify the differences in PS vs PE assembly patterns across larger experimental datasets.

**Statistical analysis of assembly patterns using fractal geometric theory:** Our first approach to statistical characterizaton of the assembly patterns was to analyze the PE and PS microparticle assemblies formed at the LC interface using fractal dimensions. To enable a fractal-based analysis of the assembly patterns, we generated a database with 355 images of PS and 491 images of PE aggregation patterns by performing 90 independent experiments in which we collected brightfield optical images of PS and PE microparticle patterns at the LC-aqueous interface across SDS concentrations of 0.01mM, 0.025mM and 0.1mM in presence of 0.3M NaCl. Each independent experiment involved placement of a TEM grid into a millifluidic channel (as shown in Figure 1a) and exposure of the TEM grid filled with LC to either PE or PS microparticles. From each TEM grid, we acquired 8-10 images, with each image showing the MP aggregation pattern in one square of the TEM grid. Our analysis included 0.025 mM SDS samples independent of whether the SDS exhibited interfacial phase separation. We used the dataset to test whether fractal dimensions (FD) permit quantitative classification of PS and PE from their assembly patterns.[71] We calculated the FD values for all the images in the microplastic image dataset via the box-counting method.[72] The FD computation scheme and FD values distribution are shown in Figure S6 and Figure 4a.

Inspection of Figure 4a reveals that the PS samples exhibited lower mean FD values with increasing SDS concentration (Figure S6): $1.38 \pm 0.10$ (0.01mM SDS), $1.22 \pm 0.05$ (0.025mM SDS) and $1.09 \pm 0.04$ (0.1mM SDS). In contrast, the FD values calculated for the PE assemblies did not change significantly with the SDS concentration: $1.32 \pm 0.04$, $1.33 \pm 0.03$, $1.33 \pm 0.04$ for 0.01mM, 0.025mM and 0.1mM SDS concentrations respectively (Figure S6c-e). Overall, these trends in the FD values are consistent with our visual analysis reported in the preceding section of this paper, namely that the organization of the PS microparticles change with SDS concentration but that the PE microparticles do not. Specifically, the PS assembly patterns change from predominantly linear chain-like structures (at 0.01 mM SDS) to a singly dispersed state (at 0.1 mM SDS)(decrease in FD), while PE samples aggregate at all SDS concentrations (constant FD) over a statistically significant dataset. We also note that the difference in FD values for PS and PE is largest at the highest SDS concentration. Next, we employed several machine learning classifiers to study the feasibility of accurately classifying the patterns as arising from PS or PE microparticles using the FD values. The performance of the machine learning (ML) architectures are shown in Figure 4b and Table S1. Inspection of Figure 4b reveals that decision tree (DT) and support-vector machine (SVM) exhibit better classification performance using our MP dataset as compared to other classifiers, with both showing higher accuracy at higher SDS concentrations as also suggested by our visual analysis. In addition, and also consistent with our visual observations, the PS and PE distributions of FDs overlap at 0.01mM SDS concentration (Figure 4a). Accordingly, the classification accuracy obtained at 0.01mM SDS concentration is only 82%. We speculate that the low classification accuracy of the geometric property-based machine learning models is because FD measures only the degree of structural symmetry in a pattern in different directions.[73,74] Specifically, patterns with different shapes can have the same fractal dimension. This limitation of FD classification led us to explore classification using deep learning, as described in the next section.

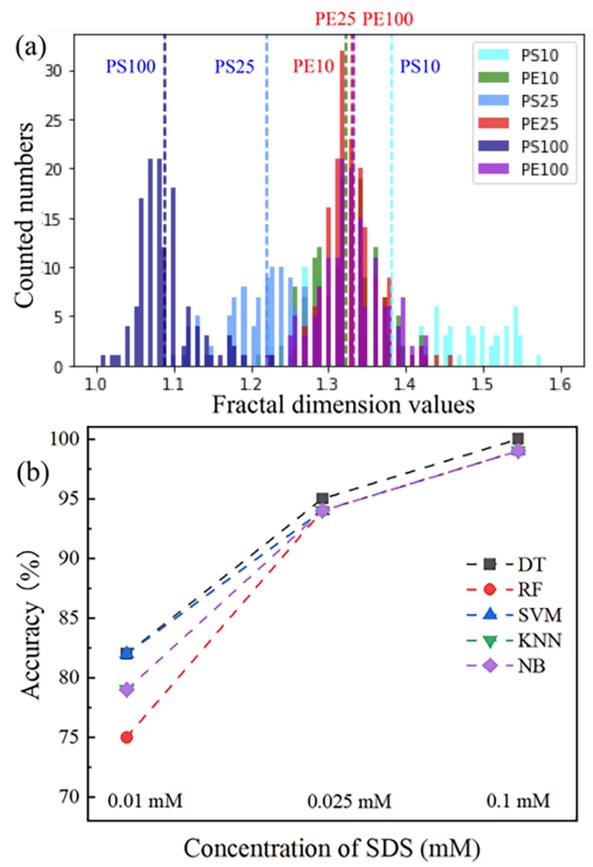

**Figure 4.** (a) Distribution of FD values calculated from all the images in the MP image dataset. The numbers placed after PS or PE indicate the concentration of SDS in μM. (b) Comparison of classification accuracy of the FD-based machine learning models over FD values. DT: decision tree; RF: random forest; SVM: support-vector machine; KNN: k-nearest neighbors; NB: naïve Bayesian.

**Deep learning-aided pattern identification:** With the goal of improving the accuracy of classification of the PE and PS assembly patterns, particularly at low SDS concentration, we developed a convolutional neural network (CNN) model. The CNN can be thought of as a combination of two components: feature extraction and classification. In contrast to the FD-based classification described in the previous section, the CNN architecture permits pixel-wise feature extraction with non-linear classification capabilities. In particular, various features of images, such as aggregate shapes, colors, and textures, can be translated into digital data and stored in feature maps.[75] This approach provides a large number of potential features by which to discriminate assembly patterns formed by either PS or PE. Among various possible CNN architectures, VGG16 has been demonstrated to be effective at feature extraction. However, the large number of parameters in the deep layer of the classification engine of VGG16 leads to slow model training[76], making it inefficient for classification, as shown in Table S3. We found it was possible to accelerate model training by separating the feature extraction and classification into two networks. Thus, we used existing VGG16 convolutional blocks as feature extractors and trained a simple CNN architecture, called "*PolyNet*", for classification (see Figure S7a). The optimized *PolyNet* hyperparameters are shown in Table S2.

We found that *PolyNet* could achieve 100% classification accuracy when trained on data obtained at a specific SDS concentration and 99.88% when measurements obtained at all SDS concentrations were used for training (Figure 5a). This represents a significant improvement in classification accuracy over the FD-ML approach (Figure 4b). We also benchmarked the *PolyNet* architecture against several alternatives (Figure S7 and Table S3 and S4) and found that *PolyNet* has the highest classification accuracy and is 35 times faster to train than VGG16. These improvements reflect a decrease in overfitting and enhanced training stability achieved by simplifying the model architecture (as described above and in Figure S7). The overall improvement of the classification achieved via use of the deep learning model (relative to FD-ML) is due to better utilization of graphic information. Each image is converted into 73728 feature vectors that store structural and textural information. This provides more features for efficient identification as compared to a single feature used in FD-ML approach. As noted above, when converting a microparticle aggregation pattern to a geometric descriptor such as FD, many features are discarded and thus FD fails to differentiate patterns when the overall geometric structures become similar.

To identify the image features that led to the improved performance of *PolyNet* relative to FD-ML, particularly when using 0.01mM SDS, we performed a feature importance analysis. This analysis identifies the parts of the input image that play an important role in the classification. For this analysis, we used a gradient-weighted class activation mapping (GradCAM). [61] This algorithm interprets the CNN classification via pixel-wise visualization. In the *PolyNet* model, we perform a so-called categorical classification that labels an image as either PS or PE using the binary vectors [0, 1] and [1, 0], respectively. By using the one-hot encoding method[77,78] followed by a softmax activation function, we obtained a final prediction score. We then determined the pixel-wise feature importance of the input image by investigating the correlation between feature maps and the final prediction score. Our CNN was trained such that larger weights are assigned to PE features to generate a final prediction score for PE. Similarly, PS features are assigned weights that correspond to the PS output score.

Using the GradCAM algorithm, we generated a qualitative color map to represent feature weights of an image classified by *PolyNet* and then superimposed it onto a binary form of the original image to visualize the regions corresponding to maximum feature importance (schematically described in Figure S8-1). The results of GradCAM analysis of *PolyNet* classification using images obtained for PS and PE microparticle patterns specifically at 0.01mM SDS are shown in Figure 5b, c (additional examples are shown in Figure S8-2). Inspection of Figure 5b reveals that the regions of the image that correlate with the presence of dense, multi-branched clusters (labelled in red) are features unique to PE, while shorter linear chains, as well as singly dispersed microparticles that are labelled red in Figure 5c, are salient PS features. Overall, the GradCAM results correlate with our visual analysis of samples in that the dense PE aggregates are assigned maximum feature importance when characterizing PE at a particular SDS concentration or across all SDS concentrations, while PS feature importance changes from short linear chains to singly dispersed microparticles from low to high SDS concentration (Figure S8-2).

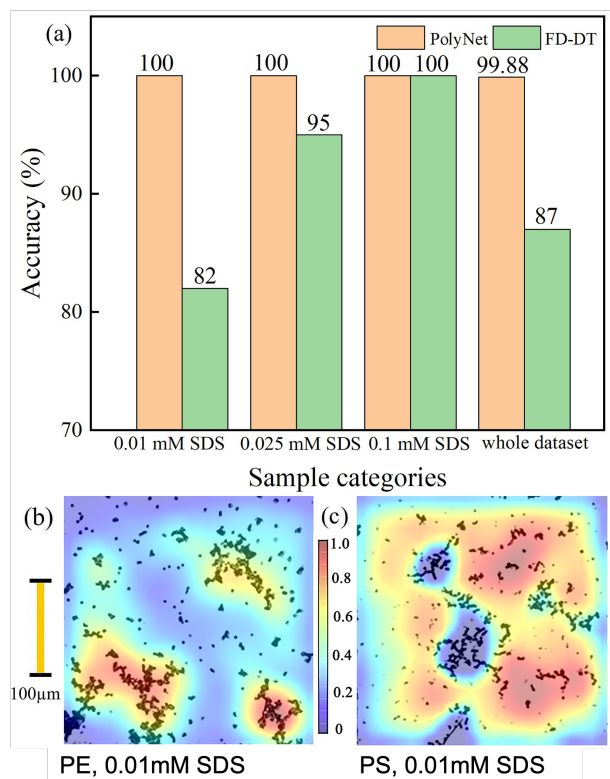

**Figure 5.** (a) Comparison plot of classification performance of our CNN model, *PolyNet* over MP image dataset versus decision tree (DT) model using FD values. (b) The GradCAM colormap of a representative PE sample at 0.01 mM SDS. (c) The GradCAM colormap of a representative PS sample at 0.01mM SDS. The colorbar from blue to red reflects feature importance from 0 to 1.

We also performed GradCAM analysis to visualize the PS features that were wrongly predicted to be PE in the overall classification (Figure S8-3). The sample (Figure S8-3a) was misclassified because parts of the image showing branched chain-like structures of PS microparticles as well as a low-resolution area at the bottom of the sample image were assigned maximum feature weights. Additionally, short or linear chains of PS that were identified to be PS features at 0.01mM SDS concentration were assigned low feature importance (Figure S8-3b). However, when trained on 0.01mM SDS only, this image was correctly classified as PS and the short and linear chains were assigned the highest feature weights (Figure S8-3c).

In summary, our results led us to conclude that *PolyNet* can classify PS and PE microparticle patterns with higher accuracy than FD-based ML methods. Decoding the CNN classification rules using GradCAM provides statistical evidence that microparticle assembly patterns formed at LC-aqueous interface by PS and PE microparticles can form the basis of highly accurate methods for identifying polymeric microparticle identity. This finding motivated us to investigate the microscopic origins of the distinct aggregation behaviors of PS and PE microparticles at the LC-aqueous interface that underlie the successful classification shown in Figure 5.

**Do differences in local LC ordering around PE and PS lead to the distinct organizations of the microparticles observed at the LC interface?** We started addressing this question by determining if LC ordering is needed to observe the distinct assembly patterns evident in Figures 2 and 3 (as used by *PolyNet* to classify the microparticle type). To this end, we

performed an experiment in which we substituted 5CB by an isotropic oil (mineral oil) (Figure S9). For PS microparticles, the chain-like PS assemblies observed at the 5CB interface were not seen at the interface of mineral oil (Figure S9-1), indicating that LC-mediated interactions underlie the PS ordering. Although the differences were less pronounced for PE microparticles, we observed differences between the PE aggregates at the interface of 5CB versus mineral oil (see Figure S9-2 for details). Additionally, we observed PE assemblies on the 5CB interface to break up when 5CB was heated across the nematic to isotropic phase transition (Figure S9-2c,d), forming patterns similar to those observed at the interface of the isotropic oil (Figure S9-2a). Overall, these results led us to conclude that the nematic ordering of the LC plays a key role in directing the PS and PE microparticles into assemblies that permit their identification.

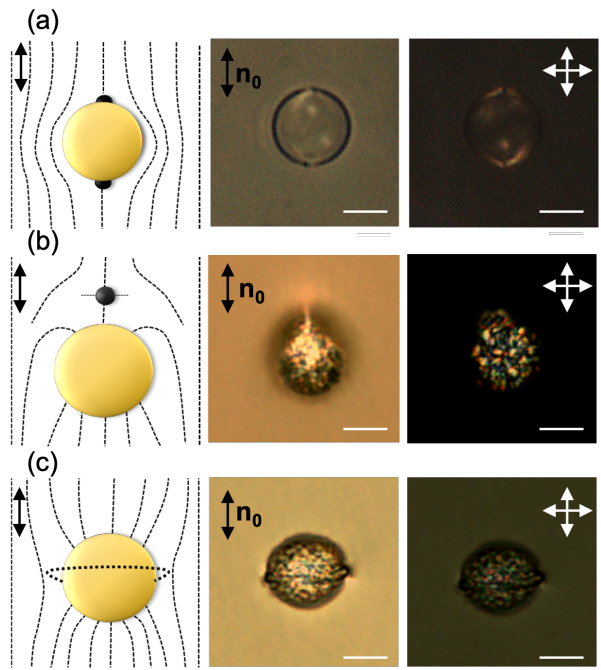

**Figure 6.** Schematic illustration of LC director profiles around a single microparticle (left column) followed by brightfield (middle column) and cross polar (right column) optical microscope images of (a) PS, boojum defect, (b) PE, hyperbolic hedgehog defect and (c) PE, Saturn ring defect in bulk 5CB, with pre-defined uniform planar anchoring on top and bottom substrates. Scale bar is 10μm.

Next, we investigated whether the assembly patterns exhibited by PS and PE microparticles at the interface of nematic 5CB could be understood in terms of local LC ordering at the microparticle surface. Past studies have reported that 5CB anchors on PS and PE microparticle surfaces with tangential[58] and perpendicular[59] orientations, respectively, conclusions that we confirmed for the microparticles used in our experiments. To facilitate microscopic observation of the ordering of the LC around the microparticles, we used PE and PS microparticles with diameters of 10-27 μm and 14-20 μm, respectively, and the same surface chemistry as the smaller polymeric microparticles used in the experiments described above. The microparticles were dispersed in bulk 5CB and observed in an optical cell (uniform planar alignment) with a thickness of 63-75 μm. We observed so-called "boojum" defects (Figure 6a) at the north and south poles of the PS microparticles, consistent with a quadrupolar configuration[79] and planar anchoring of 5CB on the PS microparticle surface (Figure 6a). In contrast, the PE microparticles exhibited either hyperbolic hedgehog defects (dipolar symmetry; Figure 6b) or Saturn ring defects (quadrupolar symmetry; Figure 6c), consistent with perpendicular anchoring of 5CB on PE surface.[80] These experiments thus confirmed that 5CB assumed a perpendicular orientation at the surface of the PE microparticles and planar orientation at the PS microparticle surfaces. Below we use this information to discuss LC-mediated interactions at the LC-aqueous interface.

Although the ordering of the LC around the PE and PS microparticles in bulk 5CB (as described above) leads us to predict that LC-mediated interparticle interactions will be either quadrupolar (planar anchoring or Saturn ring defect with radial anchoring) or dipolar (hyperbolic hedgehog defect with radial anchoring) in symmetry, if located at the LC-water interface with a contact angle of 90°, all configurations of the LC are predicted to give rise to dipolar interparticle interactions (Figure 7a and 7c for planar anchoring, and Figure 7e and 7g for radial anchoring)[52]. We note that the assumption of a contact angle of 90° was supported by experimental measurements that revealed the contact angles of the LC-aqueous interface to be $96 \pm 2°$ and $110 \pm 7°$ for PS and PE, respectively (see Figure S15). The predicted dipolar interaction at the aqueous-LC interface can be described as:

$$U_D = \frac{4\pi K C r_c^4 (1 - 3\cos^2\theta)}{L^3} \quad [2]$$

where K is the average elastic constant of the LC ($10^{-11}$ N for 5CB), C is a constant [estimated to be ~25 from theory[81] and ~6 from experimental measurements [82]], $r_c$ is the radius of the microparticle, θ is the angle between the dipoles of interacting microparticles, and L is the separation between the center of mass of the microparticles. When θ is 90°, the topological dipole is oriented along the surface normal (Figure 7a), corresponding to experiments with 0.1 mM SDS (perpendicular anchoring of the LC at the aqueous interface). Equation 2 predicts a repulsive interaction between particles for this situation, a prediction that is consistent with our experimental observations for the case of 0.1mM SDS with PS microparticles singly dispersed at the interface (Figure 2d) and low FD values (Figures 4a) [ $1.09 \pm 0.04$ ]. For lower SDS concentrations, which generate tilted anchoring of 5CB at the interface (θ < 90°), the microparticle-defect topological dipole will increasingly tilt away from the surface normal (Figure 7a,c)). This tilting of the dipole is evident in polarized optical micrographs obtained with PS microparticles at the LC-aqueous interface (Figure 7b,d). Consistent with the predictions of Eq. 2, we observe the interparticle interactions to become attractive as the LC tilts at the aqueous interface, as evidenced by microparticle chaining (Figure 2a-c) and progressively higher FD values as highlighted in Figure 4a (PS25 [$1.22 \pm 0.05$] and PS10 [$1.38 \pm 0.10$]). Overall, we conclude that the SDS-dependent organization of PS microparticles at the aqueous-LC interface can be largely understood on the basis of equilibrium predictions of LC-mediated interparticle interactions. Specifically, the organization reflects an interplay between the anchoring of 5CB at the aqueous interface and the dipolar symmetry of the LC strain/ about each microparticle.[52,83] For PE microparticles, with a contact angle close to 90° at the LC-aqueous interface, we predicted that the quadrupolar (Saturn ring) defects observed in bulk 5CB would be replaced by a single point defect (dipolar symmetry; Figure 7e, g). Consistent with this prediction, microscopic characterization of individual PE microparticles at the LC interface (Figure 7f, h) revealed an optical signature generally consistent with dipolar symmetry (see below for additional discussion of Figure 7f, h).

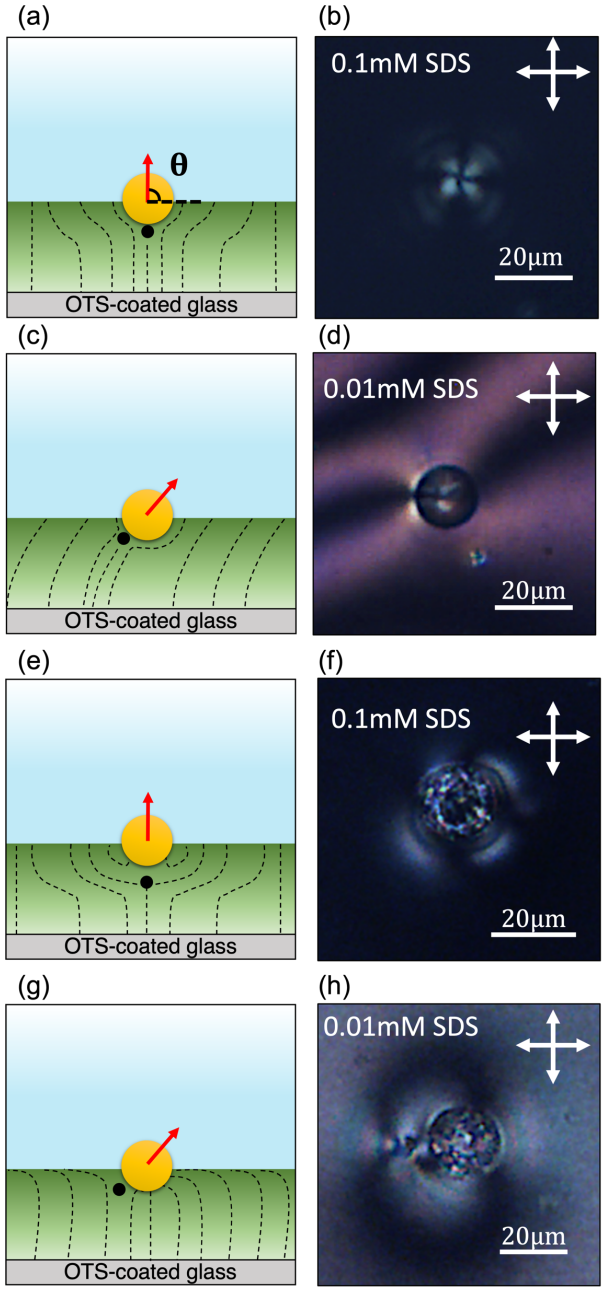

**Figure 7.** Schematic illustrations (a, c, e, g) showing side views of predicted LC director orientations (dashed lines) around a single PS (a and c) or PE (e and g) microparticle (yellow circle) at the aqueous-LC interface, when the anchoring of 5CB at the aqueous-LC interface is perpendicular (a, e) or tilted (c, g), respectively. Top view cross-polar optical micrographs of individual microparticles of PS (b and d) and PE (f and h) at the LC-aqueous interface for (b, f) 0.1mM SDS and (d, h) 0.01mM SDS, at 0.3M NaCl. Both (d) and (h) show microparticles with dipolar symmetry at the LC-aqueous interface; (d) is characterized by a boojum defect located at the pole of the PS microparticle buried in 5CB and (h) is formed by a hyperbolic hedgehog defect near a single PE microparticle. The red arrow indicates the microparticle-defect dipole.

These results led us to predict that PE microparticles would exhibit assembly patterns similar to PS under our experimental setup. Instead, however, as reported above, the assembly behavior of PE is distinctly different from PS. Specifically, across all SDS concentrations, we consistently observe PE assemblies to form a similar dense morphology (Figure 3a-d), with an average FD value of 1.33. Furthermore, as discussed above, a recurring multi-branched aggregate morphology, identified from GradCAM analysis is present in a majority of the PE samples and labelled as a feature unique to PE. We interpret the results above to indicate that PE microparticles, in contrast to PS microparticles, are assuming out-of-equilibrium (kinetically trapped) assembly states at the LC-aqueous interface. To understand why the PE microparticles do not generate equilibrium assemblies that reflect a dipolar interaction, we performed additional imaging of the PE microparticles in bulk 5CB (Figure 6b,c), in air and in mineral oil (Figure S11c,d). Inspection of Figures S11c,d reveals that, even when imaged in air or mineral oil, PE microparticles exhibit bright birefringent domains within their bulk and at their surfaces, consistent with the presence of polycrystalline high-density polyethylene (HDPE).[84] These polycrystalline domains influence the surface topography of the PE microparticles, and can be imaged (electron microscopy) as rough anisotropic features on the microparticle surface, protruding to distances of 10-70nm from the surface of the PE sphere (see SEM images of PE microparticles in Figure S11e). In contrast, the atactic PS microparticles have smooth surfaces and exhibit no observable birefringence when imaged in air or mineral oil (Figure S11a,b). Guided by these observations, we evaluated three possible reasons for the out-of-equilibrium assembly behaviour of PE microparticles at LC interface:

**1. Capillary forces generated by surface roughness:** We considered the possibility that the pinning of contact lines on the rough PE microparticle surfaces generated capillary forces acting between the microparticles on the LC-aqueous interface.[85–87] To assess this possibility, we evaluated the magnitude of the energy of interaction (W) between two capillary quadrupoles as,

$$\Delta W(L) = -12\pi\sigma H^2 \cos(2\varphi_A + 2\varphi_B)\left(\frac{r_c}{L}\right)^4 \quad [3]$$

where $H$ is amplitude of the undulation of the contact line with average radius $r_c$; $\varphi_A$ and $\varphi_B$ are the angles subtended between the diagonals of the respective quadrupoles and the line connecting the centers of the two interacting particles, defined as L, and σ is the interfacial tension of the LC-aqueous interface.

If two interacting PE microparticles are assumed to spontaneously rotate to reach an optimal orientation, the cosine function in Eq. 3 will be equal to one (maximal attractive interaction energy). Using σ=30mN/m (for 5CB-water interface), H=100nm (estimated as 10% of the microparticle radius of PE microspheres[85]), $r_c = 1\mu m$, and L = 2μm (for two microparticles in contact, L = 2$r_c$), we calculate an upper bound on ΔW of $10^4 k_B T$. This capillary interaction energy is comparable with LC-mediated interparticle interactions with energies of $\sim 10^4 K_B T$/particle (from [2], calculated using same values of $r_c$, L as in [3]). This result suggests that the organization of the PE microparticles likely reflects an interplay between capillary forces and LC-mediated interactions.

Additional support for a role for capillary forces in our results with PE is obtained from experiments performed with isotropic oil. Specifically, when comparing PS and PE microparticles at the interface of isotropic oil (Figure S9-1 and 2), PE microparticles aggregate to a greater extent: This correlates with a larger amplitude of contact line undulation, consistent with a higher roughness of the PE microparticle surface. We emphasize, however, that additional observations do suggest that LC elastic interactions also play an important role in guiding the interactions of PE microparticles. For example, the multi-branched clusters that we identified from our GradCAM analysis as characteristic of PE assembly patterns at the LC-aqueous interface, are not observed in absence of nematic order.

2. **Aggregation of PE microparticles in the bulk aqueous phase:** Next we considered the possibility that the assemblies formed by PE on the LC interface reflected pre-aggregation in the bulk aqueous phase. Microscopic observations revealed the presence of aggregates of PE (Figure S12) when preparing aqueous dispersions of PE to contact with LC. Although addition of SDS and sonication minimized the number of aggregates, they could not be avoided completely. To determine if aggregates in the bulk aqueous phase contributed to the aggregation patterns observed on the LC interface, we tracked PE particles during formation of the assemblies at the LC interface (Figure S13). We observed single PE particles to arrive at the interface, to migrate laterally on the interface to form small clusters, and to eventually form larger multi-branched aggregates. From this observation, we conclude that the dense, multi-branched clusters of PE microparticles observed on the LC interface are not a result of "pre-aggregation' in the bulk aqueous phase but rather the consequence of a growth process occurring on the interface that was fed by the arrival of single PE particles at the interface.

3. **Weak and/or non-uniform LC anchoring on the PE microparticle surfaces:** Above we proposed that the rough PE microparticle surfaces lead to large capillary forces acting between PE microparticles at the LC-aqueous interface. Here we consider the possibility that the roughness of the PE microparticle surfaces also impact LC anchoring. To evaluate this possibility, we revisited our microscopic observations of the anchoring of LC on the surfaces of the PE microparticles at the LC-aqueous interface (Figure 7f,h). Under cross polars, PE microparticles with a dipolar symmetry axis oriented along the surface normal should generate an optical image with four symmetric lobes (as observed with the PS microparticle in Figure 7b). We found, however, that many PE microparticles did not generate optical signatures comprised of four lobes; Figure 7f is an example in which only three lobes are observed (more examples of the optical signatures of PE microparticles at the LC interface are shown in Figure S14). A number of prior studies have explored the effects of surface roughness on LC anchoring, and it has been shown that surface roughness can locally disorder ("melt")[88,89] LC and cause weak anchoring. Consistent with this proposal, Kim et. al[59], measured the anchoring energy of 5CB on PE microspheres to be $10^{-6}$ Jm$^{-2}$, which is at least an order of magnitude smaller than the anchoring energy reported for PS[90], $10^{-5} - 10^{-4}$ Jm$^{-2}$. We conclude that variations in surface roughness across individual PE microparticles likely also contributes to the non-uniformity of surface anchoring discussed above.

Overall, the results above highlight the central role that microparticle surface roughness plays in determining microparticle organizations at the LC-aqueous interface. For PE and PS microparticles, the difference in surface roughness arises from the amorphous versus polycrystalline nature of the polymers within the microparticles. In particular, the high surface roughness of PE microparticles impacts the relative importance of capillary and LC elastic interactions, leading to organizations of PE microparticles at the LC-aqueous interface that reflects both classes of interactions. In contrast, the smooth surfaces of the PS microparticles leads to interfacial organizations at the LC-aqueous interface that are consistent with LC-mediated elastic interactions. Overall, our results support the proposal that LC-aqueous interfaces offer new approaches to development of surface-sensitive methods of characterization of MPs.

CONCLUSIONS

This study demonstrates the promise of LC-aqueous interfaces as the basis of surface-sensitive methods that can be used to classify the presence of polymeric microparticles with distinct surface properties. Specifically, we show it is possible to recognize two types of polymeric microparticles, PS and PE, from their unique organization patterns at LC-aqueous interfaces. We demonstrate our ability to accurately classify the assembly patterns over a statistically significant dataset using a fractal geometry-based pattern morphology analysis and a novel CNN model (*PolyNet*). *PolyNet* achieves better classification performance over FD-based machine learning for PS and PE assembly patterns at different SDS concentrations and has 35 times faster training speed as compared to baseline deep learning models and feature extraction-based ML models. Furthermore, with the help of a feature importance analysis using the GradCAM algorithm, we reveal a persistent multi-branched aggregate structure as a motif unique to PE that is largely absent in PS patterns. The PS patterns, in contrast, change with SDS concentration. We identify these features as reliable ones to classify PE from PS microparticle aggregation patterns.

Our experimental observations further revealed that the identifying assembly patterns formed by PE at LC interfaces arise from non-equilibrium phenomena whereas PS assembly patterns can be understood based on the equilibrium LC-mediated elastic interactions. We concluded that the polycrystalline nature of the PE microparticles leads to (i) stronger capillary forces than are present for PS microparticles (which have smooth surfaces and are amorphous), and (ii) weaker and less uniform LC anchoring than observed with PS microparticles. We emphasize that the differences in behaviors of the microparticles at the LC-aqueous interface reflect their surface properties, which in turn arise from amorphous versus polycrystalline polymeric microstructures.

Overall, the results of this study advance experimental and statistical methods that permit use of the LC-aqueous interface as a multifunctional interface for capture and surface-sensitive characterization of microplastics. The observations reported in this paper also generate a range of questions that warrant additional investigation. For example, we do not fully understand the relative roles of intermolecular and topographical interactions that underlie the distinct anchoring behaviors of 5CB at PE versus PS microparticle surfaces. Furthermore, in this paper, we compare the assembly patterns of microparticles formed from either PS or PE. The practical implementation of the concept proposed in this paper will require that assemblies formed by mixtures of microparticles that differ in shape, size, and chemical composition be analyzed at LC interfaces by using machine-learning approaches. In this context, we envisage that analysis of the time-dependent evolution of the assemblies formed by microparticles may provide additional information that permits complex mixtures to be classified using AI-based approaches.

**Safety Statement:** No unexpected or unusually high safety hazards were encountered.

## ASSOCIATED CONTENT

Supporting Information

The raw microplastic aggregation images can be found on the GitHub website: https://github.com/PEESEgroup/Microplastic-Project. The paper and the *Supporting Information* show all analytical and supportive results. Data and code are available from the corresponding author on request.

## AUTHOR INFORMATION

Corresponding Author(s)

Nicholas L. Abbott - Smith School of Chemical and Biomolecular Engineering, Cornell University, Ithaca, New York 14853, United States;

*E-mails: nabbott@cornell.edu

Fengqi You- Smith School of Chemical and Biomolecular Engineering, Cornell University, Ithaca, New York 14853, United States;

*E-mails: fengqi.you@cornell.edu;

Author Contributions

The experiments were performed by FM and the machine-learning was performed by AS. The manuscript was written through contributions of all authors. All authors have given approval to the final version of the manuscript.

Funding sources

This work was primarily supported by the National Science Foundation (EFMA-029327).

Notes

The authors declare no competing financial interest.

For Table of Contents Only

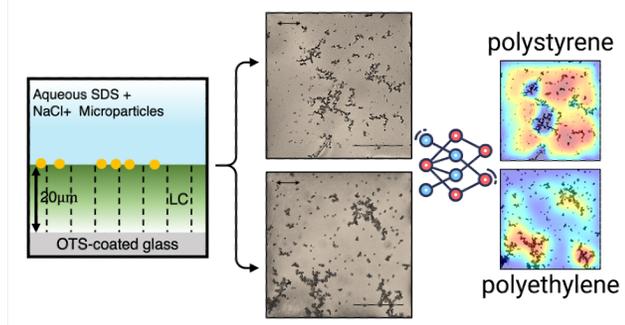